\documentclass{ckm}                 

\usepackage{mathptmx}          

\newcommand{\be}{\begin{equation}}
\newcommand{\ee}{\end{equation}}
\newcommand{\ba}{\begin{eqnarray}}
\newcommand{\ea}{\end{eqnarray}}

\newcommand{\nn}{\nonumber}

\confname{Workshop on the CKM Unitarity Triangle, IPPP Durham, April
  2003}

\title{On the resonant and non-resonant contributions to $B\rightarrow \rho \pi$}

\author{JA Oller\thanks{I would like to thank I. Bediaga for useful discussions. This contribution is partially supported by the DGICYT project FPA2002-03265.}}

\address{Departamento de F\'{\i}sica. Universidad de Murcia.\\ E-30071,
Murcia. Spain.\\
oller@um.es}

\begin{document}

\begin{abstract}
We discuss the importance of the background in order to understand the 
scattering in the $J^{\,PC}=0^{++}$ low and intermediate energy region and in particular 
regarding the $\sigma$ meson. In order to appreciate better its importance we compare with 
the $\rho$ meson in the P-wave $\pi\pi$ scattering. We also point out that in present 
analyses of three-body heavy meson decays, like those of $D^+$ and $B$, the role of this background 
 is still not properly settled  although it happens to be considerably smaller. 

\end{abstract}

\maketitle


\section{Introduction}
We refer to the talks of S. Gardner and A. Deandrea in these proceedings for a more general discussion on the 
formalism for $B\rightarrow \rho \pi\rightarrow 3\pi$. Here we just want to emphasize those aspects  more peculiar to
 the associated hadron physics through the final state interactions (FSI) that affect the $3\pi$ final state. The  aim 
 is to  measure the 
$\alpha$ CKM angle by analysing the Dalitz plot of $B\rightarrow \rho\pi\rightarrow 3\pi$ by assuming isospin invariance and 
the dominance of the $\rho$ meson through quasi two body decays \cite{56}. 

However, the influence of the broad $\sigma$ resonance in the 
$B\rightarrow 3\pi$ decays was considered  in ref.\cite{deandrea}, motivated by the E791 Collaboration \cite{e791} analysis of the $D^+\rightarrow \pi^-\pi^+\pi^+$ decay where the $\sigma$ meson was clearly seen. The chain $B\rightarrow \sigma \pi\rightarrow 3\pi$ was considered as an extra contribution to  $\rho\pi$ and then was qualified as non-resonant. However, the parameterization used in ref.\cite{deandrea}, the same as employed in ref.\cite{e791}, actually corresponds to a resonant one since the $\sigma\pi$ contribution, despite the $\sigma$ is broad, has a clear structure very different to that of a flat background as 
shown in fig.\ref{fig:pipikk}. Later on, in ref.\cite{gardner} it was pointed out that the parameterization employed in refs.\cite{e791,deandrea} gives rise to phase shifts in disagreement with the experimental ones for the isospin ($I$) 0 S-wave $\pi\pi$ partial wave \cite{kaminski}. Interestingly, the authors 
in ref.\cite{gardner} employed the parameterization of ref.\cite{meisso} to take care of the FSI induced by the $\sigma$ channel that respects chiral symmetry at low energies as well as the strong constraints from unitarity and analyticity that operate in the scalar sector and then the experimental $I=0$ S-wave $\pi\pi$ phase shifts were reproduced as well. In addition, ref.\cite{gardner} also noted that once the $\sigma\pi$ contribution was included respecting all the previous constraints, a very good reproduction of the measured ratio by CLOE \cite{12}:
\be
\label{cleo}
{\cal R}=\frac{Br(\bar{B}^0\rightarrow \rho^{\pm}\pi^{\mp})}{Br(B^-\rightarrow \rho^0 \pi^-)}=2.7\pm 1.2~,
\ee  
was obtained. A similar improvement was also noticed before in ref.\cite{deandrea}, employing the parameterization of ref.\cite{e791} to correct by FSI. Notice that this ratio of ratios is roughly 6 if one works at tree level and uses the naive factorization approximation for the hadronic matrix elements \cite{14}. 

\section{Hadron Physics}
If we consider for definiteness the chain $B^0\rightarrow \rho\pi\rightarrow 3\pi$ then one has the coherent sum of the
 amplitudes: 
$a_{+-}=A(B^0\rightarrow \rho^+\pi^-)$, $a_{-+}=A(B^0\rightarrow \rho^-\pi^+)$ and $a_{00}=A(B^0\rightarrow \rho^0\pi^0)$, multiplied by the hadronic transition amplitudes $f_+=A(\rho^+\rightarrow \pi^+\pi^0)$, $f_-=A(\rho^-\rightarrow \pi^-\pi^0)$ and 
$f_0=A(\rho^0\rightarrow \pi^+\pi^-)$, respectively. So that the following coherent sum results:
\be
\label{fa}
A(B^0\rightarrow \pi^+\pi^-\pi^0)=f_+ a_{+-}+f_- a_{-+}+ f_0 a_{00}~.
\ee
These are the commonly called resonant contributions. In addition to these contributions, in refs.\cite{deandrea,meisso} was pointed out that the non-resonant contribution $B\rightarrow \sigma \pi$, 
 which indeed is also resonant, plays a non-negligible role through the chain $ B^0(\bar{B}^0)\rightarrow \sigma \pi^0 \rightarrow \pi^+\pi^-\pi^0$ and then one needs the transition amplitude:
\be
\label{sigmap}
f_\sigma=A(\sigma\rightarrow \pi^+\pi^-)~,
\ee
so that the contribution $A(B\rightarrow \sigma \pi) f_\sigma$ should be added to eq.(\ref{fa}).

Here we want to address the issue of what does really mean the chains $B\rightarrow \rho\pi$ and $B\rightarrow \sigma \pi$, since both the $\rho$ and $\sigma$ have rather large widths, 150 MeV for the $\rho$ and $ 500$ MeV for the $\sigma$.  In order to do that we will isolate the contributions from the $\rho$ and $\sigma$ poles in $\pi\pi$ scattering and to consider the importance of these contributions compared with those coming from the background, that is, non-pole contributions.

 We accomplish this by performing a Laurent series around the pole position in the unphysical Riemann sheet\footnote{The sheets of the complex $s$-plane for one channel ($\pi\pi$) are defined with respect to the branch 
cut of the channel three-momenta $q$. The sheet I or physical corresponds to 
$\hbox{Im}q\geq 0$ and the sheet II corresponds to changing the sign of the three-momentum.} with
negative imaginary part. Due to hermitian analyticity, $f(s)=f^*(s^*)$, with $f$ a partial wave, and requires that 
poles in complex locations must occur in complex-conjugate pairs, so that a resonance is to be associated with a pair of poles on an unphysical sheet rather than a single pole \cite{spearman}.

\begin{figure}
\hbox to\hsize{\hss
\includegraphics[width=\hsize]{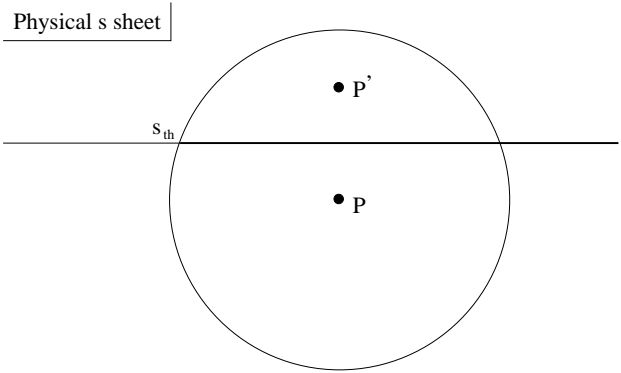}
\hss}
\caption{Circle of convergence of the Laurent series around the pole position $P$ in the II Riemann Sheet. Part of this convergence radious is located in the physical Riemann sheet.}
\label{fig:plane}
\end{figure}

Consider a resonance associated with poles at $P,$ $P'$ on the unphysical sheet (sheet II). The positions $P$, 
$P'$ are shown on fig.\ref{fig:plane}, although the diagram actually represents the physical sheet; sheet II is 
reached by crossing the cut originating from the threshold branch point $s_{th}$. It is important to realize that
 the pole at $P$ is much closer to the physical region, since we have $f(s+i\epsilon)=f_{II}(s-i\epsilon)$ 
with $\epsilon\rightarrow 0^+$ and $f_{II}(s)$ denotes the partial wave in the unphysical region.

Then we perform around $P$ a Laurent expansion of $f_{II}(s)$ which is connected continuously with
 $f$ in the physical sheet for $\hbox{Im}\,s>0$ and $\hbox{Re}\,s>s_{th}$, see 
fig.\ref{fig:plane}. Thus we have: 
\be
\label{laurant}
f_{II}(s)=\frac{\gamma_0^2}{s-s_P}+\gamma_1+\gamma_2(s-s_P)+...~,
\ee
where $\gamma_0^2$ is the residue of the pole and $\gamma_i$, $i>0$, additional coefficients  
in the series. In the table \ref{table:coup} we show the different $\gamma_0^2$ 
and $\gamma_i$ constants for the $\sigma$ and $\rho(770)$ pole positions for the
 partial waves $f_0$ and $f_1$, in order, where the subscript refers to the $\pi\pi$ angular
 momentum $\ell$. Thus, $\ell=1$ is the P-wave $I=1$ partial wave and $\ell=0$
 is the S-wave $I=0$ one. The partial waves have been taken from 
ref.\cite{nd} where they were calculated following the so called Chiral Unitary
 Approach 
\cite{nd,chupt,meisso}. This is a scheme that performs a chiral expansion 
of a softer 
interacting kernel rather than on the scattering amplitudes themselves. The point is 
to resum the unitarity cut to all orders, keeping the analyticity 
properties of this cut. This scheme can be applied both to scattering as well as to production processes, 
e.g. two photon fusion to two pseudoscalars \cite{gama}, $\phi$ radiative decays 
\cite{plb} and $J/\Psi$ decays \cite{meisso}. This approach has the advantage 
of generating both the resonant as well as the background contributions. For us, resonant 
contributions are those originating directly from the pole, the $\gamma_0^2/(s-s_P)$ 
term in eq.(\ref{laurant}), while non-resonant contributions refer to the rest of the 
terms in the Laurent expansion.

\begin{table}
\begin{center}
\begin{tabular}{|r|r|}
\hline
$\sigma$ & $\rho$ \\
\hline
 $s_\sigma=(0.466-i\,0.224)^2\,\, \hbox{ GeV}^2$ & $s_\rho=(0.758-i\,0.075)^2\,\, 
\hbox{ GeV}^2$ \\
$\gamma_0^2=5.3+i7.7$ GeV$^2$ & $\gamma_\rho^2=-5.6+i2.6$ GeV$^2$\\
$\gamma_1=-8.1+i 36.9$ & $\gamma_1=-11.3+i 1.7$ \\
$\gamma_2=1.1+i 0.1$ GeV$^{-2}$ & $\gamma_2=2.2-i1.4$ GeV$^{-2}$\\
$g_{\sigma \pi\pi}=1.4+i2.7$ GeV & $g_{\rho \pi\pi}=2.4-0.5$ GeV\\
\hline
\end{tabular}
\caption{Parameters for the Laurent expansions for the $\sigma$, first column, and 
$\rho$, second column.}
\label{table:coup} 
\end{center}  
\end{table}
In table \ref{table:coup}  we have denoted by $g_{R\pi\pi}=\sqrt{\gamma_0^2}$  and they represent the couplings of the $\sigma$ and $\rho$ to two pions. They already indicate the main difference between both resonances, that is, the much more important role played by the background in the case of the $\sigma$ 
than for the $ \rho$ meson. Let us note that the phase of $g_{\sigma\pi\pi}$ is $62.5^o$, quite large, while that of the $g_{\rho\pi\pi}$ is only  $-12^o$. Indeed, one expects the couplings of a resonance to be purely real, according to our view of elementary fields in some effective Lagrangian. This is much closer to be true for the $\rho$ case than for the $\sigma$. 
 
\begin{figure}[ht]
\hbox to\hsize{\hss
\includegraphics[width=\hsize]{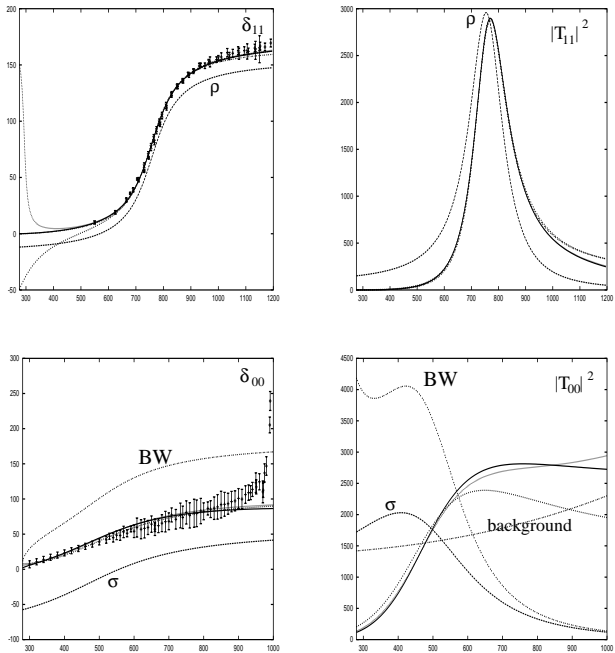}
\hss}
\caption{Different contributions according to eq.(\ref{laurant}) to the phase shifts 
and squares of the modules  of the partial waves for the $\rho(770)$ (upper panels) and 
$\sigma$ (lower ones) sectors.  The solid lines are the full results of 
ref.\cite{nd} with only the $\pi\pi$ channel. In the lower panels the lines indicated by BW  correspond to the 
relativistic Breit-Wigner parameterization, eq.(\ref{bw}). These lines are also 
present in the upper two figures for the $\rho$ channel although they are only 
barely distinguishable from the solid lines at the right end of the figures.  
In addition, the dashed lines are the 
poles contributions from the first term on the r.h.s of eq.(\ref{laurant}) and are marked with the name 
of the resonance, $\rho$ or $\sigma$. The 
short-dashed lines correspond to keep the first two terms on the r.h.s of 
eq.(\ref{laurant}) while the dotted lines corresponds to keep all the terms shown 
in the r.h.s of eq.(\ref{laurant}). We also show in the panel at the right 
lower corner as clearly indicated, the isolated total background contribution from the sum 
$\gamma_1+\gamma_2(s-s_P)^2$.}
\label{fig:pipikk}
\end{figure}

We see explicitly in the figures of the upper row of fig.\ref{fig:pipikk}, that the pole 
contribution alone is quite close to the full results for the case of the 
$\rho$ meson. Nevertheless, even in this case the differences are not negligible. 
For example, the peak of the pole contribution is shifted to lower energies than 
the peak of the full amplitude. This is also clear when looking at table 
\ref{table:coup}, since the real part of the  $\rho$ pole is 758 MeV while 
the experimental phase shifts cross $90^o$ at 770 MeV, the nominal mass for the
$\rho(770)$. 

In refs.\cite{e791,deandrea} a relativistic Breit-Wigner parameterization (BW) was 
used to mimic the influence of the broad $\sigma$ meson in the FSI of the 
$3\pi$ system through quasi-two body dominance close to the $\sigma$ pole. 
In this way the $\pi\pi$ scattering amplitudes are approximated by:
\ba
\label{bw}
T(\pi\pi\rightarrow \pi\pi)&=&
-\frac{g_{\rho\pi\pi}^2}{s-M_\rho^2+iM_\rho \Gamma_\rho(s)}~,\nn\\
T(\pi\pi\rightarrow \pi\pi)&=&
-\frac{g_{\sigma\pi\pi}^2}{s-M_\sigma^2+iM_\sigma \Gamma_\sigma(s)}~,
\ea
with $\Gamma_\sigma(s)$ and $\Gamma(s)_\rho$ energy dependent widths. 

It is also interesting then to compare the pole contribution from the $\rho$ 
with the result from the BW, eq.(\ref{bw}). We see that the BW for the $\rho$ is 
closer to the full results (solid lines) than the pure pole contribution. This is
somewhat suspicious, since the standard justification of a BW is just because it 
takes into account the pole of a nearby resonance. Indeed, through a BW one 
recovers unitarity (although no the analyticity properties associated with 
unitarity) and this maybe the reason why in the case of the $\rho$ a BW works so 
well. But we consider such a good agreement rather accidental and indeed the $\sigma$ meson 
is a perfect counter-example. 
This is clearly shown in the lower panels of fig.\ref{fig:pipikk} where the 
BW prescription is completely inadequate to reproduce the solid lines. We also 
see that the background itself is as important as the $\sigma$-pole contribution 
and the enhancement in the experimental $|T_{00}|^2$ 
\cite{munich} is a very distorted pole contribution due to the presence of 
a large background.  The distortion affects the position of the maximum as well 
as its width, as shown in the figure in the lower right corner. The message 
from this discussion is clear, whenever one is taking into account the 
S-wave $I=0$ partial wave amplitude itself or because of FSI, one should worry 
about the possible presence 
of large backgrounds that can completely modify the $\sigma-$pole shape as shown 
in the lower two panels of fig.\ref{fig:pipikk}.

Indeed, there are good theoretical reasons why there {\it should} be a strong background in the S-wave $I=0$
 $\pi\pi$ partial wave. If we think of the chiral limit, 
where the lightest quarks are massless so that the pions as well are massless, the 
interactions between the pions would vanish at $s=0$, with $s$ the 
Mandelstam variable. This is a well known result from chiral symmetry and the Goldstone 
meson nature of the pions \cite{weinberg}. However, since the mass of the pion 
is already quite low, it is then natural to admit that the $\sigma$ pole 
will remain  in the chiral limit, as indicated by some models \cite{kyoto}. These zeros are just 
required by kinematics for the higher partial waves, $\ell\geq 1$, but not for the 
S-waves. Thus, if at rather low energies one has a S-wave  pole, like the $\sigma$, then 
a strong background  that cancels this pole contributions is required in order to recuperate 
the chiral zero at $s=0$. This basic fact is not only lacking in the 
BW picture but in addition the BW parameterization used in refs.\cite{e791,deandrea} generates 
a bound state just below the $\pi\pi$ threshold, with a mass around 200 MeV. The effects of this 
bound state can 
be clearly seen in the $|T_{00}|^2$ panel of fig.\ref{fig:pipikk} since they are the responsible for the 
rapid increase of $|T_{00}|^2$ around the $\pi\pi$ threshold.

Now, if we keep in mind the previous theoretical reason for the presence of a large background, 
that is, the need to preserve the S-wave zeros as required by chiral symmetry, one can then hope that the
 background can be much smaller when no such energy dependent zeros are required. This is indeed the case for 
the $D$ and $B$ decays to $3\pi$. Here, the $D$ and $B$ mesons can be identified with pseudoscalar sources
 coupling directly to a pion while the other two pions can be thought to couple just to a scalar source. 
As a straightforward application of the Chiral Perturbation Theory (CHPT) power counting \cite{gasleut} one 
realizes that this is not suppressed by any power of momentum or quark mass and then there is a 
$priory$ no reason why the $\sigma$ meson should be screened by such huge backgrounds.\footnote{Since the 
scalar or pseudoscalar form factors are not renormalization group invariant there is a global 
factor containing light quark masses. Nevertheless, this is just a multiplicative factor in $f_\sigma$ and 
guarantees that when 
$m_q\rightarrow 0$ the amplitude for the production of pseudoscalars goes to zero independently of the parameterization 
employed. I would like to thank A. Pich for emphasizing this to me.}

\begin{figure}[ht]
\hbox to\hsize{\hss
\includegraphics[width=\hsize]{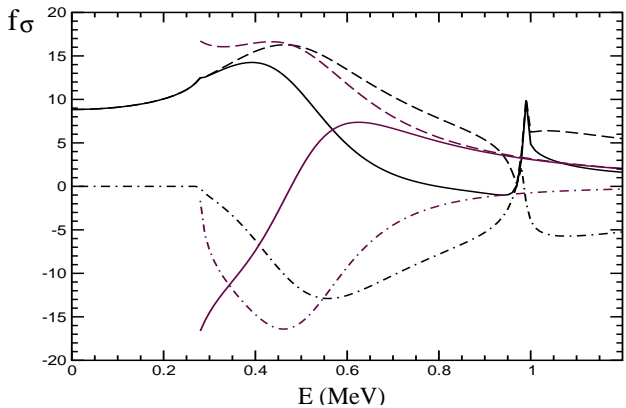}
\hss}
\caption{Transition amplitude $f_\sigma(\sigma\rightarrow \pi\pi)$ proportional to $\Gamma(s)$, as taken 
in ref.\cite{meisso}. The real and imaginary parts and absolute value of $f_\sigma$ are drawn with the 
dashed-dotted, solid and dashed lines, respectively. With the same type of curves, we also show $f_\sigma$ from the 
BW parameterization employed in refs.\cite{e791,deandrea}. These are the lines that do not continue below 
threshold.}
\label{fig:gardner}
\end{figure}

 Related to this, we consider the scalar $\pi\pi$ form factor $\Gamma(s)$ measuring the strength of 
the coupling of two pions with an $(\bar{u}u+\bar{d}d)/\sqrt{2}$ source such that:
\be
\langle 0| \frac{\bar{u}u+\bar{d}d}{\sqrt{2}}|\pi\pi \rangle=\sqrt{2}B_0\, \Gamma(s)~.
\label{ffs}
\ee 
$\Gamma(s)$, between other scalar form factors, was calculated in ref.\cite{meisso} such that 
an algebraic matching with the CHPT series is obtained at low energies and at the same time 
full unitarity and analyticity are kept to all orders in the chiral expansion. 
This form factor is proportional (with an expected soft $s$ dependence for this 
proportionality constant) to $f_\sigma(\sigma\rightarrow \pi\pi)$) \cite{gardner}. The latter is shown 
in fig.\ref{fig:gardner}, where we also compare with the corresponding $f_\sigma$ from the BW 
parameterization used in refs.\cite{e791,deandrea}.

We see that the real and imaginary parts are particularly different between the results from 
ref.\cite{meisso,gardner} and the simpler BW parameterization \cite{e791,deandrea}, however the absolute value is 
quite close between both solutions. This is in sharp contrast to what we have shown in fig.\ref{fig:pipikk} 
for the two lower panels since there both the pole and BW results are in complete disagreement with the full result, 
solid lines. This indicates that since the scalar form factor lacks of chiral zeros, there is then 
no theoretical reason for a huge background and 
the $\sigma$ meson peak is much better seen. Indeed, we have explicitly checked this point 
by performing the Laurent expansion of $\Gamma(s)$ of ref.\cite{meisso}. This can be the reason why 
in $B$ and in particular 
in $D$ decays \cite{e791}, one is clearly seeing the $\sigma$ meson although this state is not so clear 
in $\pi\pi$ scattering, as discussed above when considering figs.\ref{fig:pipikk} and fig.\ref{fig:gardner}. 
Notice that  $f_\sigma$ of ref.\cite{gardner} satisfies the constraint that its phase 
coincides with the phase shifts of S-wave $I=0$ $\pi\pi$ scattering, since it is proportional to $\Gamma(s)$ 
 from ref.\cite{meisso} that fulfills unitarity with a $T-$matrix whose phase shifts agree well with 
the experimental ones. This constraint is expected to 
occur in good approximation for $B$ and $D$ decays to $3\pi$  when two of the pions are around the $\sigma$ meson 
mass with its quantum numbers.  Then, the resulting $\sigma \pi$ system is very energetic and 
the strong interactions between them should be 
rather  soft and cannot be responsible for a large deviation in the phase of $f_\sigma$ with respect to 
the $\pi\pi$ $I=0$ S-wave phase shifts, as expected from a naive application of the Watson's theorem.

To sum up, we have shown that in general terms there is no pole dominance associated with the $\sigma$ 
meson as explicitly shown for $\pi\pi$ scattering. We have considered this fact as a consequence of the 
presence of the chiral zeros in S-waves. We have then argued that when these are absent it could well happen
 that the $\sigma$ meson pole 
dominates with just a  slight background.

\end{document}